\documentclass[epj]{svjour}
\usepackage{graphics}
\begin{document}
\title{Search for the onset of baryon anomaly at RHIC-PHENIX}
\author{Tatsuya Chujo for the PHENIX Collaboration
}                     % Do not remove
\offprints{chujo@sakura.cc.tsukuba.ac.jp}   % Insert a name or remove this line
\institute{Institute of Physics, University of Tsukuba, Tsukuba, Ibaraki 305-8571, Japan}
\date{Received: date / Revised version: date}
\abstract{
The baryon production mechanism at the intermediate $p_T$ (2 - 5 GeV/$c$) at RHIC is still not well 
understood. The beam energy scan data in Cu+Cu and Au+Au systems at RHIC may provide us a further 
insight on the origin of the baryon anomaly and its evolution as a function of $\sqrt{s_{NN}}$. In 2005 
RHIC physics program, the PHENIX experiment accumulated the first intensive low beam energy data 
in Cu+Cu collisions. We present the preliminary results of identified charged hadron spectra in Cu+Cu 
at $\sqrt{s_{NN}}$ = 22.5 and 62.4 GeV using the PHENIX detector. The centrality and beam energy 
dependences of (anti)proton to pion ratios and the nuclear modification factors for charged pions and 
(anti)protons are presented.
\PACS{
      {25.75.Dw}{} %\and
%      {PACS-key}{discribing text of that key}
     } % end of PACS codes
} %end of abstract
\maketitle
%
% ===================================================================
% ================
% 1. Introduction
% ================
\section{Introduction}
\label{sec:intro}

One of the most surprising observations at RHIC is a particle type dependence of
hadron yield suppression at the intermediate transverse momentum $p_T$ (2 - 5 GeV/$c$)~\cite{PPG026,PPG015}.  
In Au+Au central collisions at $\sqrt{s_{NN}} = 200 $ GeV, yields for mesons are largely suppressed ~\cite{PPG014} 
with respect to the yields in proton-proton collisions in the intermediate $p_T$, while those for baryons 
are not suppressed, and show a binary nucleon-nucleon collision ($N_{coll}$) scaling. 

In d+Au experiment at RHIC, there is also a particle type dependence in the nuclear modification 
factor $R_{AA}$~\cite{PPG030}. The $R_{AA}$ for protons in d+Au is larger than that for pions and kaons.
This observations can be understood by Cronin effect~\cite{cronin1975,antreasyan1979}. 
It is found, however, that the particle species dependence in d+Au is not large enough to account 
for the absence of suppression for protons as seen in Au+Au central collisions at RHIC. This phenomena 
in Au+Au central collisions is 
called "Baryon Anomaly at RHIC". To explain the baryon anomaly, many theoretical models have 
been proposed, such as quark recombination models~\cite{recombi}, 
%a baryon junction model~\cite{junction},
and a hybrid model of hydrodynamics with a color glass condensate (CGC) and jet quenching~\cite{hydro}. 
All of these models are able to reproduce the experimental data qualitatively for both pions and protons. 

On the elliptic flow ($v_2$) measurements, there is a also clear particle species dependence~\cite{PPG022}. 
The number of constituent quark scaling of $v_2$ shows a clear universal curve 
regardless of the particle species, and these scaling properties support the quark recombination 
picture for the hadron production at RHIC. 

In order to test the quark recombination picture, $\phi$ meson is one of the ideal test 
particles, because the mass is similar to protons', but it's a meson particle. If the hydrodynamical 
collective flow is a dominant source of baryon anomaly, $\phi$ meson's $R_{AA}$ should
follow the data for protons (no suppression) due to its heavy mass. 
Thanks to the high statistics data in Au+Au collisions at $\sqrt{s_{NN}} = $ 200 GeV taken in 2004 
RHIC run, it is confirmed that the $R_{AA}$ for $\phi$ mesons behaves like a meson~\cite{phi_raa}, 
and its $v_2$ follows the same universal quark number scaling curve of all other particle species~\cite{phi_v2}.
Giving the fact that $\phi$ mesons behave like a "meson", it is now widely believed
that the quark recombination is one of the main mechanisms for the hadronization 
process at the intermediate $p_T$ in central Au+Au collisions at RHIC. 

Now the key questions are; where the onset of the baryon anomaly is, and how they evolve 
as a function of beam energy. In order to answer these questions, the lower energy beam
data in Cu+Cu collisions were taken during the 2005 RHIC run by the PHENIX experiment. 
In this paper, we present the preliminary results of single particle $p_T$ spectra 
for identified charged particles in Cu+Cu collisions at $\sqrt{s_{NN}}$ = 22.5 GeV and 62.4 GeV
measured by the PHENIX experiment. The beam energy and centrality dependences of
the particle ratios and nuclear modification factors are presented and discussed.

% =================
% 2. Data analysis
% =================
\section{PID charged particle spectra analysis}
\label{sec:ana}

% data set
The low beam energy data sets in Cu+Cu have been taken during the 2005 RHIC run. 
We analyzed 5.2 M minimum bias events for 22.5 GeV Cu+Cu, and 66.5 M 
minimum bias events for 62.4 GeV Cu+Cu data set. We constructed the tracks 
within a collision vertex $\pm$ 30~cm from the center of the spectrometer. 

% detector & tracking, PID 
Charged particle tracks are reconstructed by the Drift Chamber (DC) based on a 
combinatorial Hough transform. The first layer of Pad Chamber (PC1) is used to 
measure the position of the hit in the longitudinal direction (along the beam axis). 
When combined with the location of the collision vertex along the beam axis (from the 
Beam-Beam Counter(BBC)), the PC1 hit gives the polar angle of the track. Only tracks 
with valid information from both the DC and PC1 are used in the analysis. 
In order to associate a track with a hit on the high resolution Time-of-Flight 
detector (TOF), the track is projected to its expected hit location on the TOF.
Tracks are required to have a hit on the TOF within $\pm$2$\sigma$ of the expected 
hit location in both the azimuthal and beam directions. The charged particle 
identification (PID) is performed by using the combination of three measurements: 
time-of-flight from the BBC and the TOF, momentum from the DC, and flight path
length from the collision vertex point to the hit position on the TOF. 

%---------
\begin{figure}
\resizebox{0.5\textwidth}{!}{
  \includegraphics{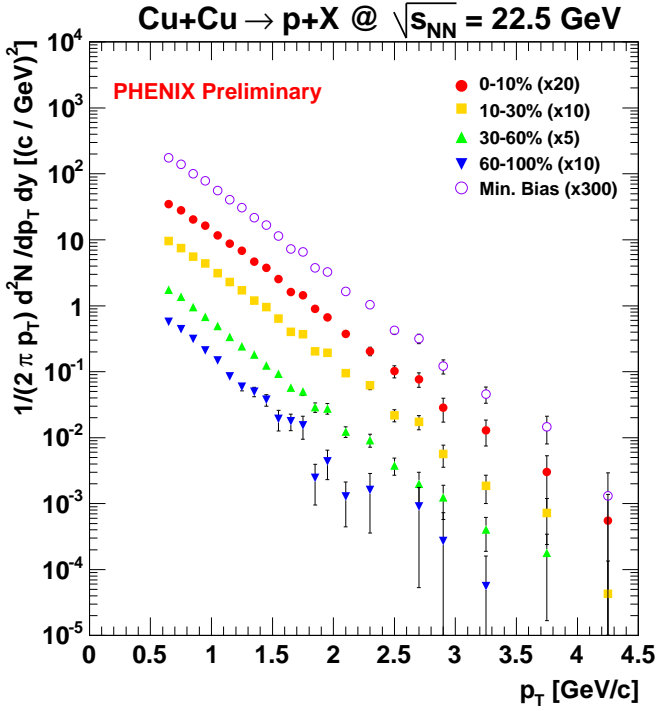}
}
\caption{Centrality dependence of $p_T$ spectra for protons
in Cu+Cu collisions at $\sqrt{s_{NN}} =$ 22.5 GeV. No feeddown correction is
applied.}
\label{fig:pt_pr_22gev} 
\end{figure}
%---------

%---------
\begin{figure}
\resizebox{0.5\textwidth}{!}{
  \includegraphics{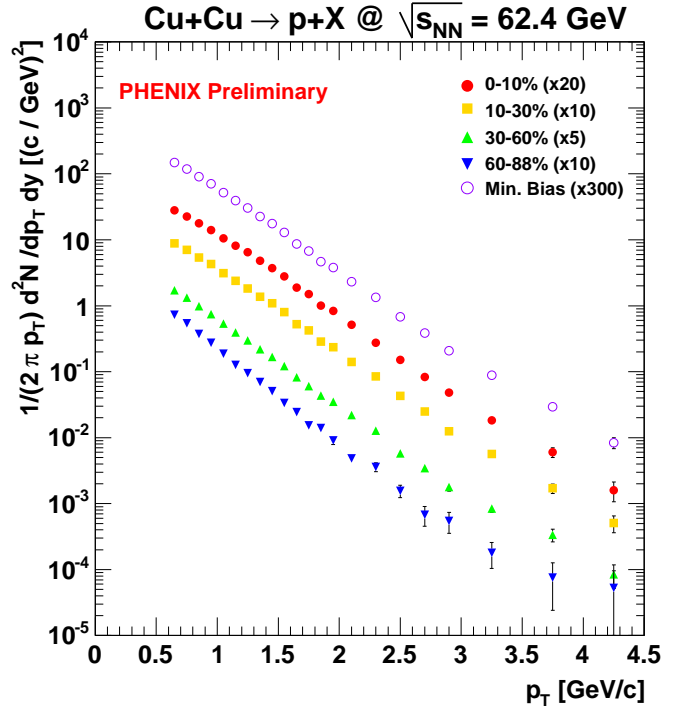}
}
\caption{Centrality dependence of $p_T$ spectra for protons
in Cu+Cu collisions at $\sqrt{s_{NN}} =$ 62.4 GeV. No feeddown correction is
applied.}
\label{fig:pt_pr_62gev} 
\end{figure}
%---------

% correction
The geometrical acceptance and the in-flight decay are corrected by
the GEANT based Monte Carlo simulation.
No occupancy correction (detector inefficiency correction due to 
the high track densities) is applied. The expected occupancy correction 
is below 4\% in Cu+Cu collisions 62.4 GeV at the most central events and less for peripheral 
collisions, the correction is negligible. No weak decay feeddown 
correction is applied for proton and antiproton spectra in this analysis. 

% centrality
The centrality determination in Cu+Cu collisions 22.5 GeV is purely based on the 
PC1 hit multiplicity in order to avoid an effect of the spectator 
nucleons. Since at $\sqrt{s_{NN}} =$ 22 GeV the beam rapidity gap is quite 
narrower than those at the higher beam energy, the BBC (located 
in 3.0~$<$~$|\eta|$~$<$~3.9) is able to see the spectator nucleons. The 
PC1 multiplicity distribution with the BBC $z$ vertex correction is 
subdivided into four centrality bins. For Cu+Cu 62.4 GeV data, we use the 
charge distribution measured by the BBC. We obtain the number of binary
collisions ($N_{coll}$) and the number of participant nucleons ($N_{part}$)
for each data set and each centrality bin by the Glauber Monte Carlo 
calculation in PHENIX.

% ==============
% 3. Results
% ==============
\section{Results}
\label{sec:result}
We present the following preliminary results 
in Cu+Cu collisions at $\sqrt{s_{NN}} =$ 22.5 GeV and 62.4 GeV, measured by the PHENIX experiment.
(1) the identified charged particle $p_T$ spectra, 
(2) particle ratios ($p/\pi^+$, $\overline{p}/\pi^-$) as a function of $p_T$ for each centrality class, and 
(3) the nuclear modification factor $R_{AA}$ for pions and (anti)protons as a function of $p_T$.

% ------------------
%  Spectra
%  -----------------
\subsection{$p_T$ spectra}
\begin{figure}
\resizebox{0.5\textwidth}{!}{
  \includegraphics{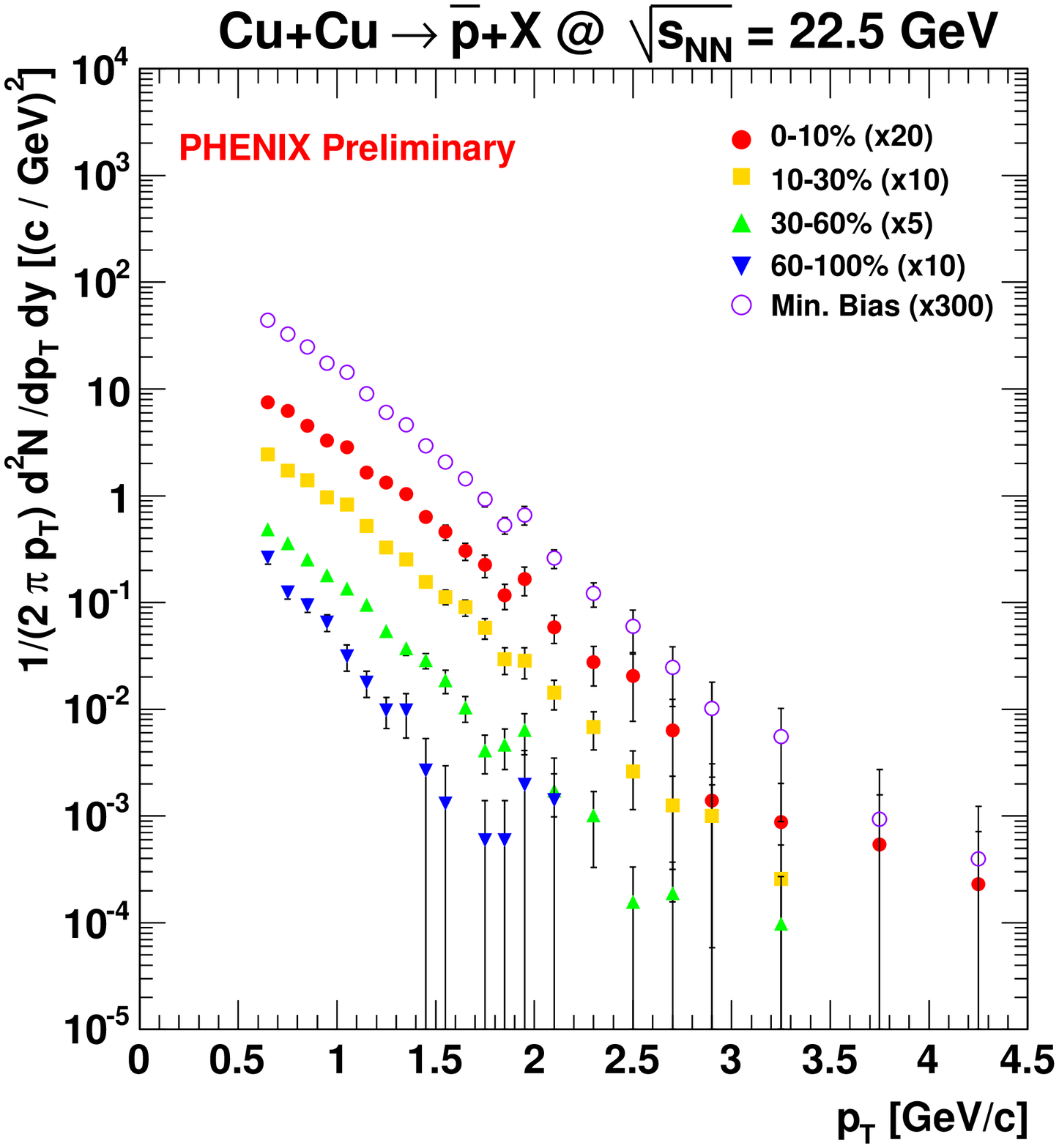}
}
\caption{Centrality dependence of $p_T$ spectra for $\overline{p}$ 
in Cu+Cu collisions at $\sqrt{s_{NN}} =$ 22.5 GeV. No feeddown correction is
applied.}
\label{fig:pt_pb_22gev} 
\end{figure}

\begin{figure}
\resizebox{0.5\textwidth}{!}{
  \includegraphics{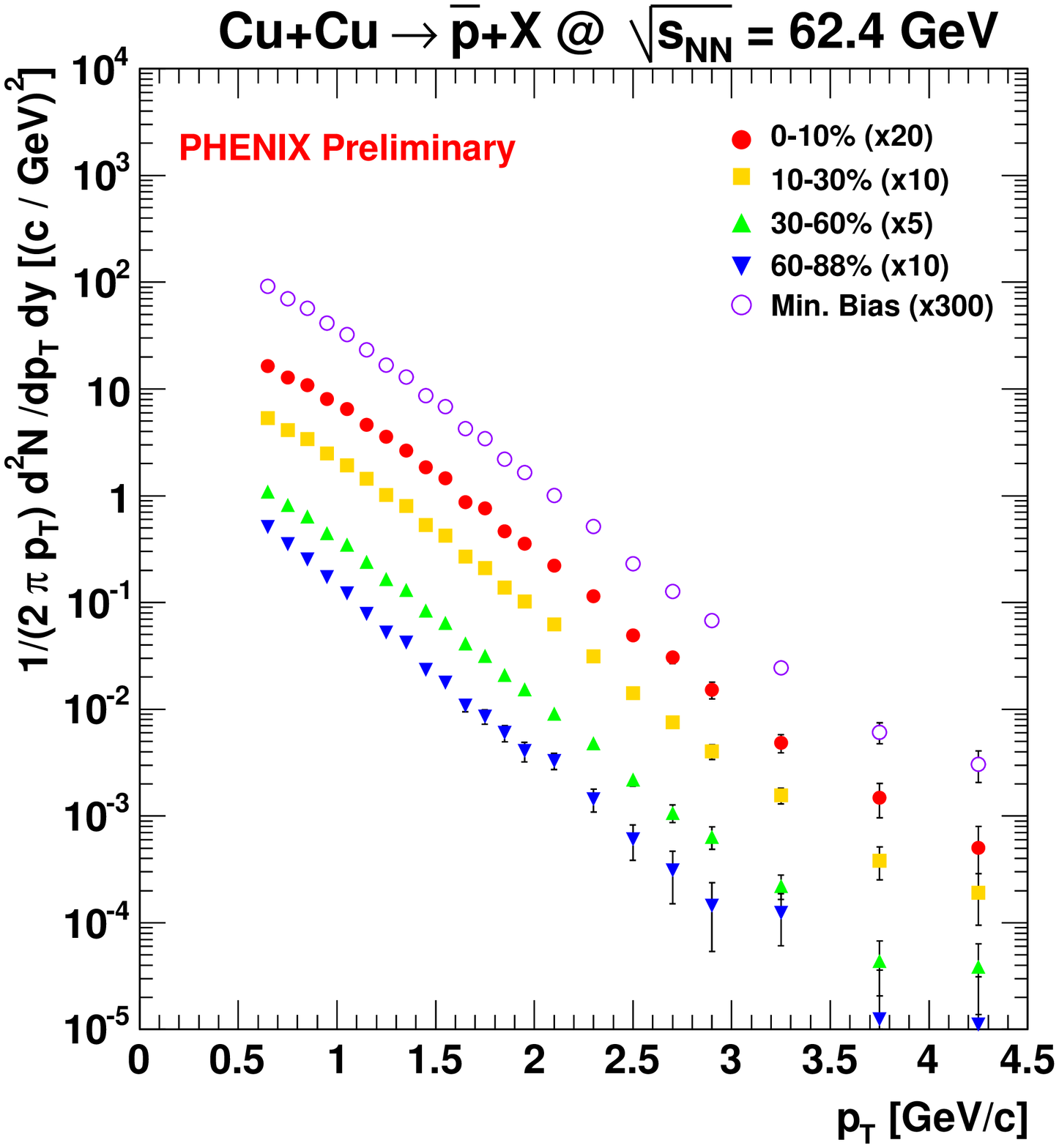}
}
\caption{Centrality dependence of $p_T$ spectra for $\overline{p}$ 
in Cu+Cu collisions at $\sqrt{s_{NN}} =$ 62.4 GeV. No feeddown correction is
applied.}
\label{fig:pt_pb_62gev} 
\end{figure}

Fig.~\ref{fig:pt_pr_22gev} and  Fig.~\ref{fig:pt_pr_62gev} show the 
centrality dependences of $p_T$ spectra for protons in Cu+Cu 
collisions at $\sqrt{s_{NN}} =$ 22.5 GeV and 62.4 GeV, respectively.
The $p_T$ spectra for antiprotons are shown in Fig.~\ref{fig:pt_pb_22gev} 
and \ref{fig:pt_pb_62gev}. No weak decay feeddown correction is applied 
for both protons and antiprotons spectra and both beam energies. 
We measured the spectra in 
four centrality bins (0-10\%, 10-30\%, 30-60\%, 60-88 or 60-100\%) 
and in the minimum bias events. The $p_T$ spectra for each centrality 
bin for $\pi^{\pm}$ and $K^{\pm}$ are also measured at both beam
energies (see~\cite{HQ_presen}). 

%---------
%---------

%---------
%---------

% ------------------
%  Particle Ratio 
%  -----------------
\subsection{Particle ratios ($p/\pi^+$ and $\overline{p}/\pi^-$)}
\begin{figure}
\resizebox{0.5\textwidth}{!}{
  \includegraphics{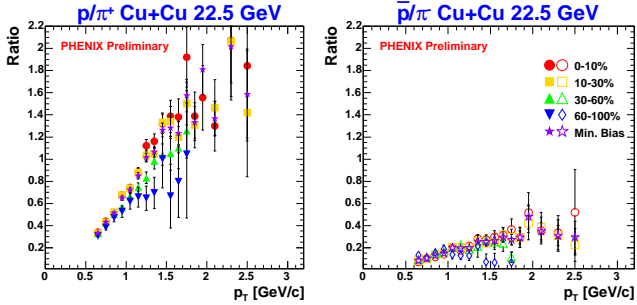}
}
\caption{Centrality dependence of $p/\pi^{+}$ (left) and $\overline{p}/\pi^{-}$
(right) ratios as a function of $p_T$ in Cu+Cu collisions at $\sqrt{s_{NN}} =$ 22.5 GeV.
No feeddown correction is applied.}
\label{fig:ppi_ratio_all_22GeV} 
\end{figure}

\begin{figure}
\resizebox{0.5\textwidth}{!}{
  \includegraphics{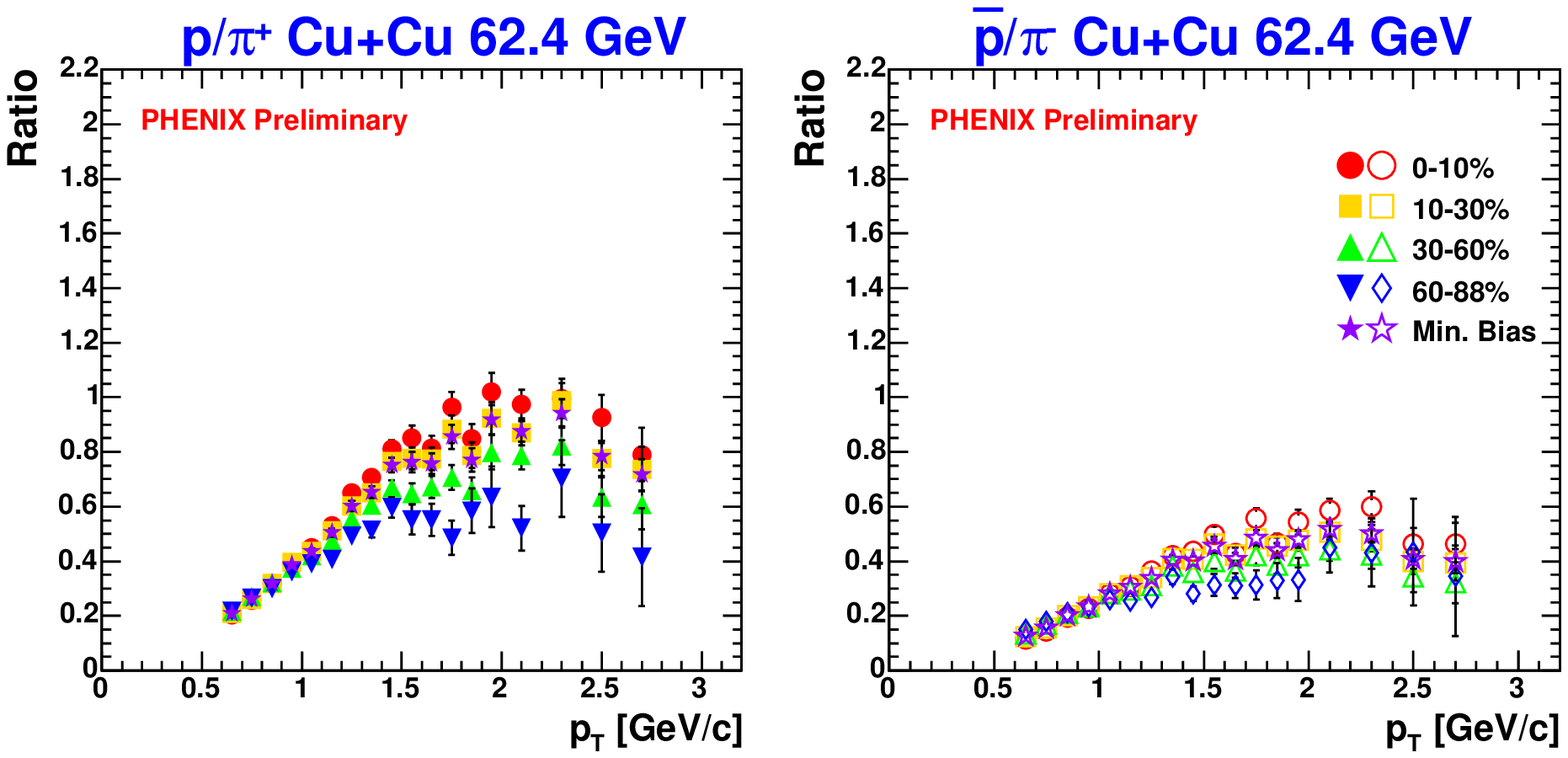}
}
\caption{Centrality dependence of $p/\pi^{+}$ (left) and $\overline{p}/\pi^{-}$
(right) ratios as a function of $p_T$ in Cu+Cu collisions at $\sqrt{s_{NN}} =$ 62.4 GeV.
No feeddown correction is applied.}
\label{fig:ppi_ratio_all_62GeV} 
\end{figure}

Fig.~\ref{fig:ppi_ratio_all_22GeV} shows the centrality dependence of
$p/\pi^{+}$ and $\overline{p}/\pi^{-}$ ratios as a function of 
$p_T$ in Cu+Cu collisions at $\sqrt{s_{NN}} =$ 22.5 GeV. The similar
plots for Cu+Cu at 62.4 GeV are shown in Fig.~\ref{fig:ppi_ratio_all_62GeV}. 

In 22.5 GeV Cu+Cu data, $p/\pi^{+}$ ratio is significantly larger than those 
in the higher beam energy in Au+Au~\cite{PPG026}, and it is increasing as a 
function of collision centrality. This large $p/\pi^{+}$ ratios can be 
understood by the influence of the incoming nucleons from the beams. 
For $\overline{p}/\pi^{-}$ ratios, (almost) no centrality dependence is observed, 
and the ratios are constant at the value of 0.3 - 0.4 at $p_T = 2.0$ GeV/$c$, 
which is close to the expected values in the fragmentation and the data in p+p 
collisions. 

In 62.4 GeV Cu+Cu data, the influence of incoming protons is smaller than
the one in Cu+Cu collisions at 22.5 GeV, as seen in $p/\pi^{+}$ ratios in Fig.~\ref{fig:ppi_ratio_all_62GeV}.
The centrality dependence of $\overline{p}/\pi^{-}$ is seen. In the most central collisions,  
$\overline{p}/\pi^{-}$ ratio is about 0.6 at $p_T = 2$  GeV/$c$, and in the peripheral 
collisions, the value becomes consistent with that for p+p collisions. 

\begin{figure*}
\resizebox{0.9\textwidth}{!}{
  \includegraphics{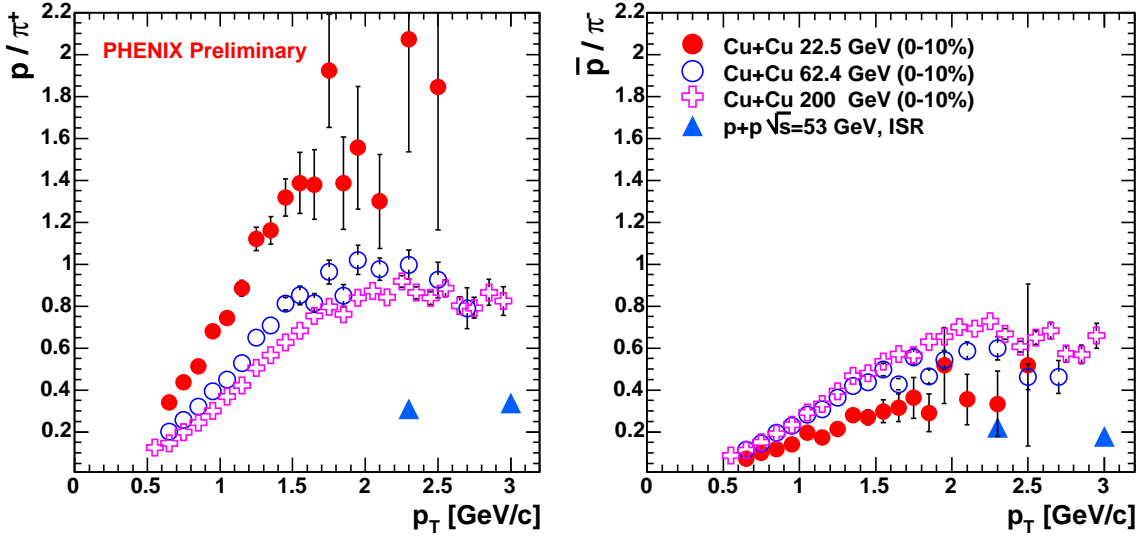}
}
\caption{Beam energy dependence of $p/\pi^{+}$ (left) and $\overline{p}/\pi^{-}$
(right) ratios as a function of $p_T$ in Cu+Cu collisions from $\sqrt{s_{NN}} =$ 22.5 GeV
to 200 GeV. The data for 200 GeV Cu+Cu is taken from the PHENIX preliminary 
data~\cite{konno_CuCu_200GeV}.}
\label{fig:ppi_ratio_sys_dep}
\end{figure*}

Fig.~\ref{fig:ppi_ratio_sys_dep} shows the beam energy dependence of $p/\pi^{+}$ 
and $\overline{p}/\pi^{-}$ ratios in Cu+Cu collisions from $\sqrt{s_{NN}} =$ 22.5 GeV
to 200 GeV~\cite{konno_CuCu_200GeV}. There is a clear beam energy dependence in Cu+Cu 
from 22.5 GeV to 200 GeV, i.e. $p/\pi^{+}$ ($\overline{p}/\pi^{-}$) decreases (increases) 
as a function of $\sqrt{s_{NN}}$, respectively. 

For the study of the baryon anomaly at the lower beam energies at RHIC, 
$\overline{p}/\pi^{-}$ ratios would be a good measure, because antiprotons are ``produced 
particles'', but measured protons are the mixture of produced particles and the incoming protons
from the beams. The absence of centrality dependence in $\overline{p}/\pi^{-}$  ratio in
Cu+Cu 22.5 GeV may suggest that there is no baryon anomaly at this beam energy. To conclude 
this observation, the high statistics data with the heavier collisions system like Au+Au 
at around 22.5 GeV is necessary in the future RHIC run, and the results should compare 
with the existing data in Pb+Pb collisions at the top SPS energy ($\sqrt{s_{NN}} = $ 17.3 GeV) 
at high $p_T$~\cite{NA49}.

%---------
%---------

%---------
%---------

%---------
%---------

% ------------------
%  R_AA
%  -----------------
\subsection{Nuclear modification factor: $R_{AA}$}

To obtain the nuclear modification factors at the lower energy, we used the $p_T$ spectra at mid-rapidity 
in p+p collisions at $\sqrt{s} = 23 $ GeV and 63 GeV for kaons and (anti)protons as the reference spectra, 
measured by the British-Scandinavian Collaboration~\cite{Alper75}. Those data are fitted by 
the empirical functional form (see~\cite{HQ_presen} in detail). For pions, we use the 
parameterization as described in~\cite{david-pp-62GeV}. 
The nuclear modification factor $R_{AA}$ is defined as the following equation:

\begin{equation} 
R_{AA}(p_T)\,=\,\frac{(1/N^{evt}_{AA})\,d^2N_{AA}/dp_T dy}{\langle N_{coll}\rangle/\sigma_{pp}^{inel} \,\times\, 
d^2\sigma_{pp}/dp_T dy},
\label{eq:R_AA}
\end{equation}
where the $\langle N_{coll}\rangle/\sigma_{pp}^{inel}$ is 
the average Glauber nuclear overlap function, $\langle T_{AuAu} \rangle$.

\begin{figure}
\resizebox{0.5\textwidth}{!}{
  \includegraphics{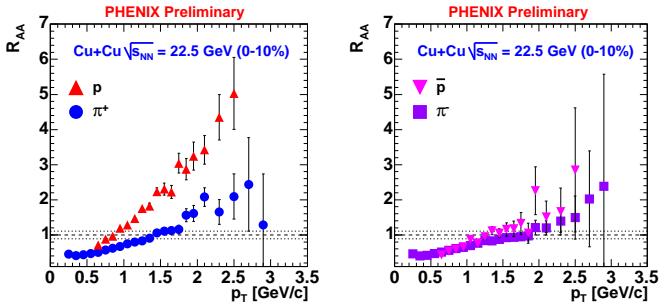}
}
\caption{Nuclear modification factor $R_{AA}$ (0-10\%) for charged pions and (anti)protons 
in Cu+Cu collisions at $\sqrt{s_{NN}} =$ 22.5 GeV. The positively charged particles are shown in the left
panel, and the negatively charged particles are shown in the right panel.
}
\label{fig:raa_all_22gev} 
\end{figure}

\begin{figure}
\resizebox{0.5\textwidth}{!}{
  \includegraphics{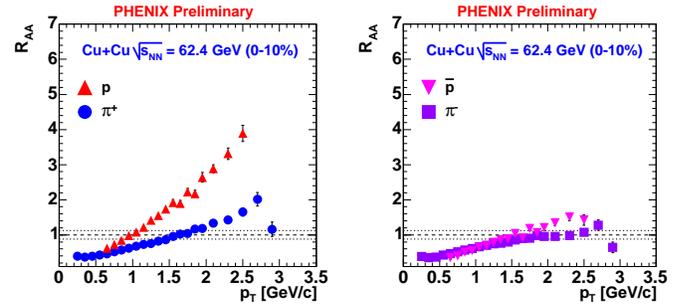}
}
\caption{Nuclear modification factor $R_{AA}$ (0-10\%) for charged pions and (anti)protons 
in Cu+Cu collisions at $\sqrt{s_{NN}} =$ 62.4 GeV. The positively charged particles are shown in the left
panel, and the negatively charged particles are shown in the right panel.}
\label{fig:raa_all_62gev} 
\end{figure}

The $R_{AA}$ for pions and (anti)protons are shown in Fig.~\ref{fig:raa_all_22gev} for 
22.5 GeV Cu+Cu, and in Fig.~\ref{fig:raa_all_62gev} for 62.4 GeV Cu+Cu. The centrality
selection for $R_{AA}$ is 0-10\% central collisions for both beam energies. The values 
of the average number of collisions used here are $\langle N_{coll} \rangle = $140.7 ($\sigma_{inel} = 32.2$ mb) 
for 22.5 GeV Cu+Cu, and $\langle N_{coll} \rangle = $152.3 ($\sigma_{inel} = 35.6$ mb) 
for 62.4 GeV Cu+Cu.
The systematic error on $R_{AA}$ consists of; (1) $<N_{coll}>$ uncertainty (one $\sigma$
error as shown in dotted-lines above and below $R_{AA} = 1$), (2) combined systematic errors
from p+p reference data and Cu+Cu spectra (not shown in the figures). 
The statistical errors are shown in the error bars on each data point.

The data show that there is no suppression on charged pions for both 22.5 and 62.4 GeV
Cu+Cu central collisions. For proton's $R_{AA}$, a clear enhancement is seen and 
can be attribute to the effect of the incoming beam nucleons. The value of 
proton's $R_{AA}$ at 62.4 GeV is slightly smaller than that at 22.5 GeV. 
$R_{AA}$ for antiprotons is almost unity at the intermediate $p_T$ for 
both 22.5 and 62.4 GeV, and is very similar to the one for pions. 

In 2006, the PHENIX experiment has successfully collected the new reference p+p 
data at $\sqrt{s_{NN}} = $ 62.4 GeV. The precision of $R_{AA}$ measurements
at 62.4 GeV for both Cu+Cu and Au+Au is expected to be improved by using 
this new p+p data set.

% ==============
% 4.Conclusions
% ==============
\section{Conclusions}
\label{sec:conclusion}

In summary, we have measured $p_T$ spectra for $\pi^{\pm}$, $K^{\pm}$,
$p$, $\overline{p}$ in Cu+Cu collisions at $\sqrt{s_{NN}} =$ 22.5 and 62.4 GeV. 

In 22.5 GeV Cu+Cu data, 
we observed a larger $p/\pi^{+}$ ratio compared to those at the higher beam
energy in Au+Au, increasing as a function of collision centrality. 
For $\overline{p}/\pi^{-}$ ratio, almost no centrality dependence is seen, 
and the ratio is about 0.3 - 0.4 at $p_T = 2.0$ GeV/$c$, which is close to 
the value in p+p collisions. In 62.4 GeV Cu+Cu data, $p/\pi^{+}$ ratios
are reduced, compared to those in Cu+Cu 22.5 GeV, and the centrality dependence
of $\overline{p}/\pi^{-}$ is seen. There is a clear beam energy dependence
on those ratios in Cu+Cu from 22.5 GeV to 200 GeV, i.e. $p/\pi^{+}$ ($\overline{p}/\pi^{-}$) 
decreases (increases) as a function of $\sqrt{s_{NN}}$, respectively. 
The observed $\overline{p}/\pi^{-}$ ratio may suggest there is no baryon 
anomaly at 22.5 GeV Cu+Cu collisions. To make this point clearer, a further 
data set (a high statistics low beam energy data in heavy collision system) 
is necessary in the future RHIC run. 

For the $R_{AA}$, no suppression is observed for charged pions for both 22.5 and 62.4 GeV
in Cu+Cu central collisions. For protons $R_{AA}$, there is a clear enhancement,
which can be attributed to the incoming beam nucleons. The $R_{AA}$ in 62.4 GeV 
is slightly smaller than that in 22 GeV. The antiprotons $R_{AA}$ is almost unity 
at the intermediate $p_T$ region for both beam energies.

%---------
%---------

%---------
%---------

% ============
%  REFERENCES  
% ============


\begin{thebibliography}{}

\bibitem{PPG026} S.~S.~Adler {\it et al.} (PHENIX Collaboration), Phys. Rev. ~C \textbf{69}, (2004) 034909.
\bibitem{PPG015} S.~S.~Adler {\it et al.} (PHENIX Collaboration), Phys. Rev. Lett. \textbf{91}, (2003) 172301.
\bibitem{PPG014} S.~S.~Adler {\it et al.} (PHENIX Collaboration), Phys. Rev. Lett. \textbf{91}, (2003) 072301.
\bibitem{PPG030} S.~S.~Adler {\it et al.} (PHENIX Collaboration), nucl-ex/0603010.
\bibitem{cronin1975} J.~Cronin~{\it et al.}, Phys. Rev. ~D \textbf{11}, (1975) 3105.
\bibitem{antreasyan1979} D.~Antreasyan~{\it et al.}, Phys. Rev. ~D \textbf{19}, (1979) 764.
\bibitem{recombi} R.~C.~Hwa and C.~B.~Yang, Phys. Rev. ~C \textbf{67}, (2003) 034902;
 R.~J.~Fries, B.~M\"uller, C.~Nonaka and S.~A.~Bass, Phys. Rev. Lett. \textbf{90}, (2003) 202303;
 V.~Greco, C.~M.~Ko and P.~L\'evai, Phys. Rev. Lett. \textbf{90}, (2003) 202302.
%\bibitem{junction} I.~Vitev and M.~Gyulassy, Nucl. Phys. \textbf{A715}, (2003), 779c.
\bibitem{hydro} T.~Hirano, Y.~Nara, Phys. Rev. ~C \textbf{69}, (2004) 034908, 
 T.~Hirano, Y.~Nara, Nucl. Phys. \textbf{A743}, (2004) 305.
\bibitem{PPG022} S.~S.~Adler {\it et al.} (PHENIX Collaboration), Phys. Rev. Lett. \textbf{91}, (2003) 182301.
\bibitem{phi_raa} D.Pal (PHENIX Collaboration),hep-ex/0510020.
\bibitem{phi_v2} M.~Issah, A.~Taranenko (PHENIX Collaboration), nucl-ex/0604011.
\bibitem{HQ_presen} HQ 2006 presentation file can be found in; \\ 
http://hq2006.bnl.gov/program.html
\bibitem{NA49} C.~Alt {\it et al.} (NA49 Collaboration), Nucl. Phys. \textbf{A774} (2006) 473.
\bibitem{konno_CuCu_200GeV} M.~Konno (PHENIX Collaboration), nucl-ex/0510022.
\bibitem{Alper75} B.~Alper {\it et al.}, Nucl. Phys.~B \textbf{100}, (1975) 237.
\bibitem{david-pp-62GeV} D.~d'Enterria, nucl-ex/0411049.

\end{thebibliography}
\end{document}